\documentclass[%
reprint,
nofootinbib,
 amsmath,amssymb,
aps,
pra,
superscriptaddress,
showkeys,
longbibliography
]{revtex4-1}

\usepackage{enumitem}
\usepackage{tabularx}
\usepackage{graphicx}

\usepackage[breaklinks, hidelinks, colorlinks=true,linkcolor=blue,citecolor=black]{hyperref}% add hypertext capabilities
\usepackage{color, colortbl}
\usepackage[table]{xcolor}
\usepackage{layouts}
\usepackage{booktabs}

\makeatletter
\renewcommand\frontmatter@abstractwidth{\dimexpr0.9\textwidth\relax}
\makeatother

\makeatletter
\newcommand*{\addFileDependency}[1]{
  \typeout{(#1)}
  \@addtofilelist{#1}
  \IfFileExists{#1}{}{\typeout{No file #1.}}
}
\makeatother

\newcommand{\sm}{\scalebox{0.5}[1.0]{\( - \)}}

\makeatletter
\renewcommand\subparagraph{\@startsection{subparagraph}{5}{\parindent}%
    {3.25ex \@plus1ex \@minus .2ex}%
    {-1em}%
    {\normalfont\normalsize\bfseries}}
\makeatother

\begin{document}

\preprint{1}

\title{Investigating the Visual Cues of CNNs for Vascular Segmentation: A Case Study in Microscopy and Fundus Imaging}

\author{Weslley dos Santos Silva}
\affiliation{Department of Computer Science, Federal University of S\~ao Carlos, S\~ao Carlos, SP, Brazil}

\author{Cesar H. Comin}
\email[Corresponding author: ]{comin@ufscar.br}
\affiliation{Department of Computer Science, Federal University of S\~ao Carlos, S\~ao Carlos, SP, Brazil}

\date{\today}

\begin{abstract}
Vascular segmentation is a standard procedure for clinical diagnosis, yet the specific visual features determining model decisions remain poorly understood. This paper investigates the visual cues Convolutional Neural Networks (CNNs) use to segment blood vessels across two distinct imaging domains: fluorescence microscopy and retinal fundus photography. We employ a series of experiments to quantify the influence of shape, texture, and receptive field on segmentation performance.
First, we isolate texture and intensity by evaluating performance on patches subjected to pixel shuffling and normalization. Second, we assess global shape relevance by training models on sparse contours and centerlines. Lastly, we quantify the required spatial context by systematically varying the network's theoretical and effective receptive fields.
Within the scope of the evaluated datasets, we found that pixel intensity is more relevant than texture, though networks maintain surprisingly high accuracy even when both cues are removed. Furthermore, CNNs struggle to extrapolate full vessel geometry from shape cues alone, typically relying on a relatively small effective receptive field of around 20 pixels, though global context provides a modest benefit for fundus images. While specific to the modalities studied, this methodology offers a quantitative foundation to audit and refine deep learning systems in vascular imaging.
\end{abstract}


\maketitle
\thispagestyle{plain}

\section{Introduction}
\label{sec:introduction}

Blood vessel segmentation is essential for analyzing conditions ranging from diabetic retinopathy to neurodegenerative diseases~\cite{mookiah2021review,chen2023all,li2022human}. However, the task is uniquely challenging due to the large morphological diversity of the vasculature. Vessels exhibit a complex combination of distinct features: elongated, branching geometries (global shape), specific internal intensity profiles (local texture), and varying contrast against background tissue. For instance, while major arteries in fundus images are clearly defined, fine capillaries in noisy microscopy samples are often hard to identify. Understanding how computational models weigh these competing visual cues is important for ensuring that models remain robust in clinical and preclinical applications.

While Convolutional Neural Networks (CNNs) have achieved state-of-the-art accuracy in these tasks~\cite{isensee2021nnu,isensee2024nnu,ma2024segment}, they are largely treated as black boxes. This lack of transparency is a significant hurdle for clinical integration, as it complicates model debugging and hinders the ability to ensure robustness against data variations. Critically, for regulatory approval and physician trust, there is a need to move beyond measuring if a model works and begin quantifying how it works~\cite{rudin2019stop,nazir2023survey}. For a vascular clinician, knowing whether a model identifies a vessel based on its anatomical continuity (shape) or merely its local pixel intensity is important for predicting where that model might fail in a pathological case.

To address this need, we look toward the \emph{shape versus texture} bias discussed in the broader computer vision community. Research on natural images has shown that CNNs often prioritize local texture over global object shape, a strategy that differs fundamentally from human perception~\cite{geirhos2018imagenet}. In the context of blood vessels, which are defined by their elongated geometry, this bias is particularly relevant. This reliance on local features is further complicated by the Effective Receptive Field (ERF) of the network~\cite{luo2016understanding}, which determines how much spatial context the model actually utilizes. If a model’s ERF is too small to capture the branching nature of the vascular tree, it may be forced to rely on local texture, potentially making it fragile to imaging artifacts. Therefore, disentangling the influence of shape, texture, and context on vessel segmentation is critical for designing more efficient, robust, and reliable systems.

To achieve this, we conducted a three-part investigation designed to systematically disentangle the influence of texture, shape, and network receptive field on segmentation performance within two distinct imaging contexts: fluorescence microscopy (VessMAP) and retinal fundus photography (DRIVE). Our methodology employs these datasets as case studies to isolate and quantify the importance of different visual cues through a series of targeted experiments. First, to probe the role of texture and intensity, we evaluated the impact of removing absolute intensity values and spatial relationships via normalization and pixel shuffling. This provides a controlled environment to quantify how performance degrades when specific information channels are restricted. Second, we investigated the capacity of representative CNN architectures to segment vessels using only sparse shape cues provided by outer contours and centerlines, thereby testing the models' ability to extrapolate vascular structures without internal appearance information. Finally, we examined the spatial context required for accurate segmentation by systematically varying the models' theoretical receptive field as well as by dividing the input images into non-overlapping patches and training a segmentation model on the patches.

Within the specific scope of our experiments, the results indicate that while pixel intensity is a more critical feature than local texture, the evaluated CNNs maintain high performance even when both cues are significantly perturbed. Moreover, the findings suggest that, for these modalities, models are unable to reliably recover full vessel geometry from sparse shape cues alone. We also observed that the effective receptive field utilized by the networks converges to approximately 20 pixels for the tasks considered, although global information provides a measurable, if modest, benefit for the more structured fundus images. While restricted to a subset of datasets and architectures, these analyses serve as a diagnostic template, providing a quantitative foundation for understanding the decision-making process of CNNs and offering a methodology for auditing the robustness of vascular segmentation models.

The remainder of the text is structured as follows. In Section~\ref{sec:related}, we provide an overview of the related work. In Section~\ref{s:methodology}, we present the datasets used in the analyses as well as a detailed description of the three aforementioned experiments. The results of each experiment are presented in Section~\ref{s:results}. In Section~\ref{s:conclusion}, we provide the conclusion of the study.

\section{Related Works}
\label{sec:related}

The search for architectural innovation for performance gains has yielded increasingly complex and opaque models that can accurately identify where a vessel is, but offer little insight into why a given pixel was classified as a vessel. This lack of clarity represents a barrier to the clinical translation of technologies and motivates a shift in focus from performance alone to performance coupled with interpretability~\cite{chaddad2023survey,nazir2023survey,singh2020explainable}. The work of Geirhos et al.~\cite{geirhos2018imagenet} showed that standard CNNs exhibit a strong bias toward local texture for object recognition. The results spurred new research into quantifying and manipulating this bias~\cite{dai2022rethinking,heinert2024reducing,tripathi2023edges}, often using style transfer to create cue conflicts or training on stylized datasets to force models to rely on shape. New studies~\cite{chung2022shape,islam2021shape,hermann2020} argued that an optimal model strikes a task-dependent balance between shape and texture sensitivity. This is particularly relevant to medical imaging, where both the shape and texture of the vasculature carry critical diagnostic information.

The body of work most relevant to our research focuses on isolating each cue to verify its influence on model performance. Dai et al.~\cite{dai2022rethinking} proposed a texture removal approach using a mean shift filter~\cite{fukunaga1975estimation}, a shape perturbation procedure that blurs object borders and a topology removal technique implemented as image patch shuffling. However, none of these proposed techniques completely removes the desired cues. For instance, a mean shift filter can still blur object borders if they are not particularly sharp. Similar procedures were adopted by Heinert et al.~\cite{heinert2025shape}.

Mütze et al.~\cite{mutze2025influence} proposed cue decompositions that are related to those defined in our work. For texture cues, a Voronoi diagram is created, and each cell receives a texture generated from patches of a single semantic class. For shape cues, they apply the Holistically-Nested Edge Detection (HED) method~\cite{xie2015holistically} to detect object edges, which are then used as input to train the network. However, the proposed methods are not applicable to blood vessels, as they are thin structures with smooth borders. Our study generates the cues using more direct manipulations: pixel normalization and/or shuffling to remove intensity and texture cues, and a segmentation-from-shape task to probe segmentation performance when there is no texture information.

A CNN's decision-making is also influenced by its receptive field (RF), the input region influencing its output. Critically, research has shown that CNNs have an ERF that is substantially smaller than the theoretical RF~\cite{luo2016understanding,zhang2024convolutional}. In the context of medical images, some results indicate a strong relationship between RF size and model performance~\cite{behboodi2020receptive,sytwu2022understanding}. 

Loos et al.~\cite{loos2024demystifying} conducted a comprehensive analysis closely related to our work. They varied the depth and kernel sizes of a U-Net model to systematically control the RF and evaluated the model on two synthetic and six real-world medical datasets. They found that the RF size is more relevant for datasets that have a low contrast between the foreground and background classes, and that diminishing returns are observed for RFs larger than the segmented objects. A key difference in our analysis is that we focus on a semantic segmentation task where vessels lack a predefined global structure. Furthermore, we include dilated convolutions in our model variations, a popular approach for increasing context while keeping the number of parameters fixed.

Vessel segmentation embodies two main aspects: it requires a large RF to guarantee vessel continuity while maintaining good local precision to delineate thin boundaries. Architectural features like dilated convolutions and downsampling help manage a trade-off between the two aspects, but a systematic analysis of the RF's role in vessel segmentation is lacking. We systematically vary the RF through architectural changes in the model, as well as the amount of context available from the input data. This allows us to empirically derive a context-performance curve of a U-Net model.

To our knowledge, no previous work has conducted a systematic experimental investigation designed to disentangle and quantify the influences of shape, texture, and network receptive field on the performance of CNNs for blood vessel segmentation. Our paper is designed to fill this gap by conducting a series of targeted experiments. 

\section{Methodology}
\label{s:methodology}

\subsection{Base datasets}

The experiments used the VessMAP~\cite{viana2025new} and DRIVE~\cite{StaalDRIVE} datasets. The VessMAP dataset comprises 100 fluorescence microscopy images of the mouse cortex, each with a resolution of $256\times 256$ pixels. The images include blood vessels exhibiting various features. Notable variations involve samples with differing levels of noise, contrast, and vessel size, along with samples presenting imaging artifacts and notable intensity fluctuations within blood vessel segments.

To determine whether the methodology is applicable to vessel imagery from different domains, we also included the DRIVE dataset. This dataset is widely used for testing vascular segmentation algorithms and comprises 20 training and 20 testing fundus images, each measuring $584 \times 565$ pixels. 

The microscopy images in the VessMAP dataset are significantly different from the fundus images in the DRIVE dataset. In the DRIVE dataset, thick vessels generally have a higher contrast against the background than those in the VessMAP dataset. Because DRIVE images capture the entire retina, they exhibit a predefined global structure that includes the optic disk and retinal borders. In contrast, VessMAP samples are acquired from very small regions of the cortex and therefore lack this global anatomical structure.

\subsection{The relevance of texture and intensity information}

To quantify the importance of textural and intensity information, we developed a classification task in which a neural network must identify whether image patches contain blood vessels. We considered two scenarios: i) patches containing strictly vessel or background textures, and ii) patches containing both, thereby including vessel borders, which can serve as important classification cues. We then selectively removed texture and intensity information from these patches to measure the resulting impact on classification accuracy.

\subsubsection*{Classification without border information}

Because blood vessels are thin structures, we selected a patch size of $9\times 9$ pixels to ensure the patches could fit entirely inside the vessels. This size provided a sufficiently large context window while still allowing the generation of an adequate number of patches for training. Larger patch sizes generated too few patches for experimentation. The identification of possible patches for extraction was done by applying a convolution operation to the ground truth images using a $9\times 9$ filter filled with ones. The pixels containing the value 81 in the output represent candidate center points for patch extraction.

From these candidate pixels, up to seven were randomly selected for each sample. To avoid overlapping patches, once a pixel was selected, all pixels in a $9\times 9$ area around it were removed from the set of candidate points. For some samples, fewer than seven patches could be extracted using the aforementioned criteria. The procedure generated 539 vessel texture samples from the VessMAP dataset. The DRIVE dataset was not used in this experiment because most of the vessels were too thin for patch extraction.

\begin{figure}[htbp]
    \centering
    \includegraphics[width=\columnwidth]{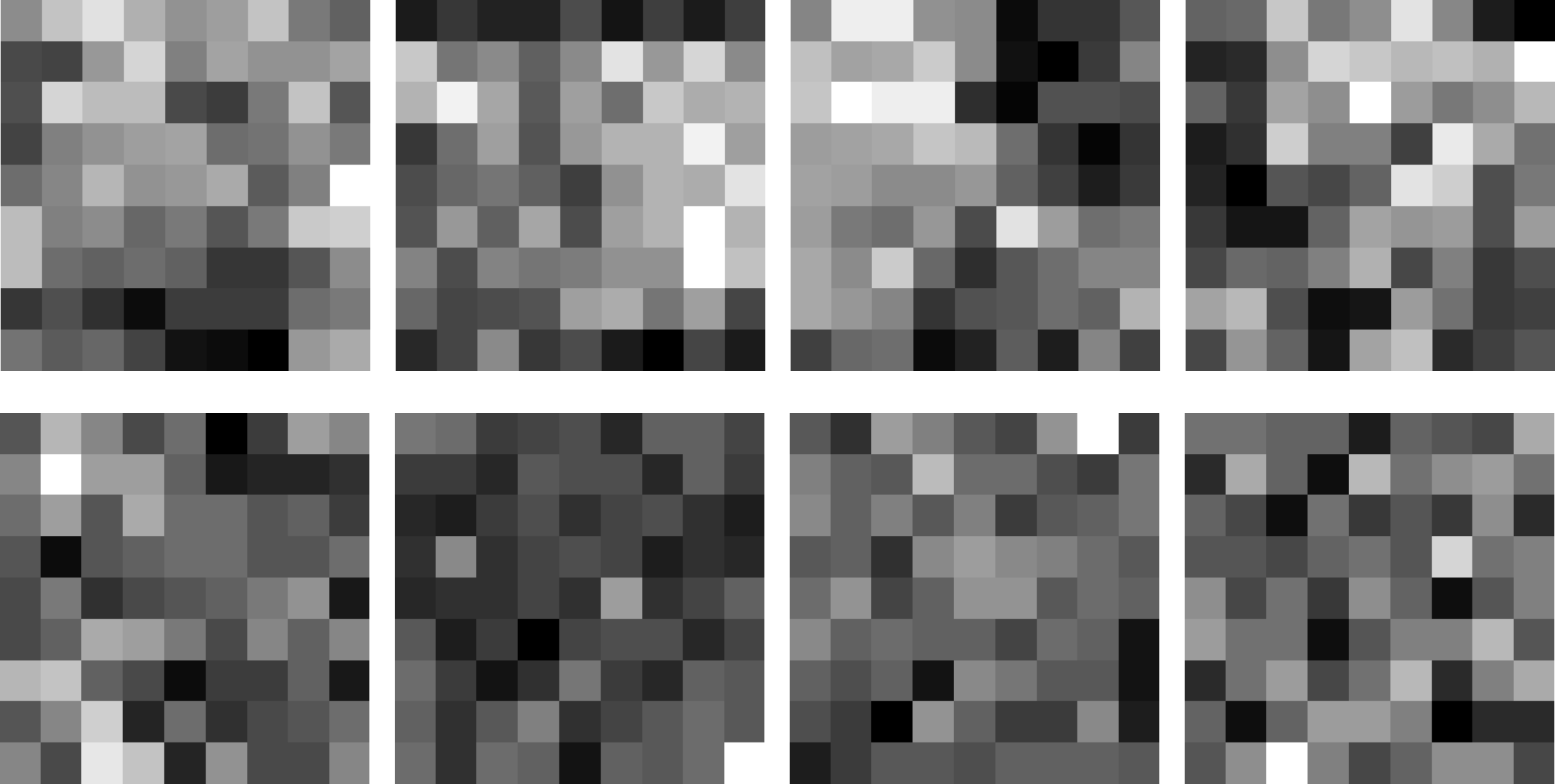}
    \caption{Examples of $9\times 9$ patches extracted from the VessMAP dataset. The first and second rows of images show, respectively, patches extracted from the blood vessels and the background of the samples.}
    \label{f:patchs_from_vessel_background}
\end{figure}

The same criteria were applied to extract background patches. A total of 700 background patches were extracted, from which 539 were randomly selected to create a balanced dataset with respect to the vessel patches. Figure~\ref{f:patchs_from_vessel_background} shows examples of extracted patches. Visually distinguishing whether a patch belongs to a vessel or the background is difficult. This challenge motivates our quantitative analysis of a neural network's ability to classify the patches.

\subsubsection*{Classification with border information}

Classifying from texture alone can be a challenging task. The addition of border cues may significantly improve classification performance as it allows contrasting the texture of the vessel with its surrounding background. To quantify the impact of border cues on classification performance, a set of patches was generated containing a specific percentage $f$ of vessel pixels. This was done using a similar procedure described above. The ground truth image was convolved with a $9\times 9$ filter filled with ones, and pixels in the output with a value of $\text{round}(81\times f/100)$ were used as candidate center pixels for patch extraction. The remainder of the procedure was the same as in the previous experiment. 

\begin{figure}[htbp]
    \centering
    \includegraphics[width=\columnwidth]{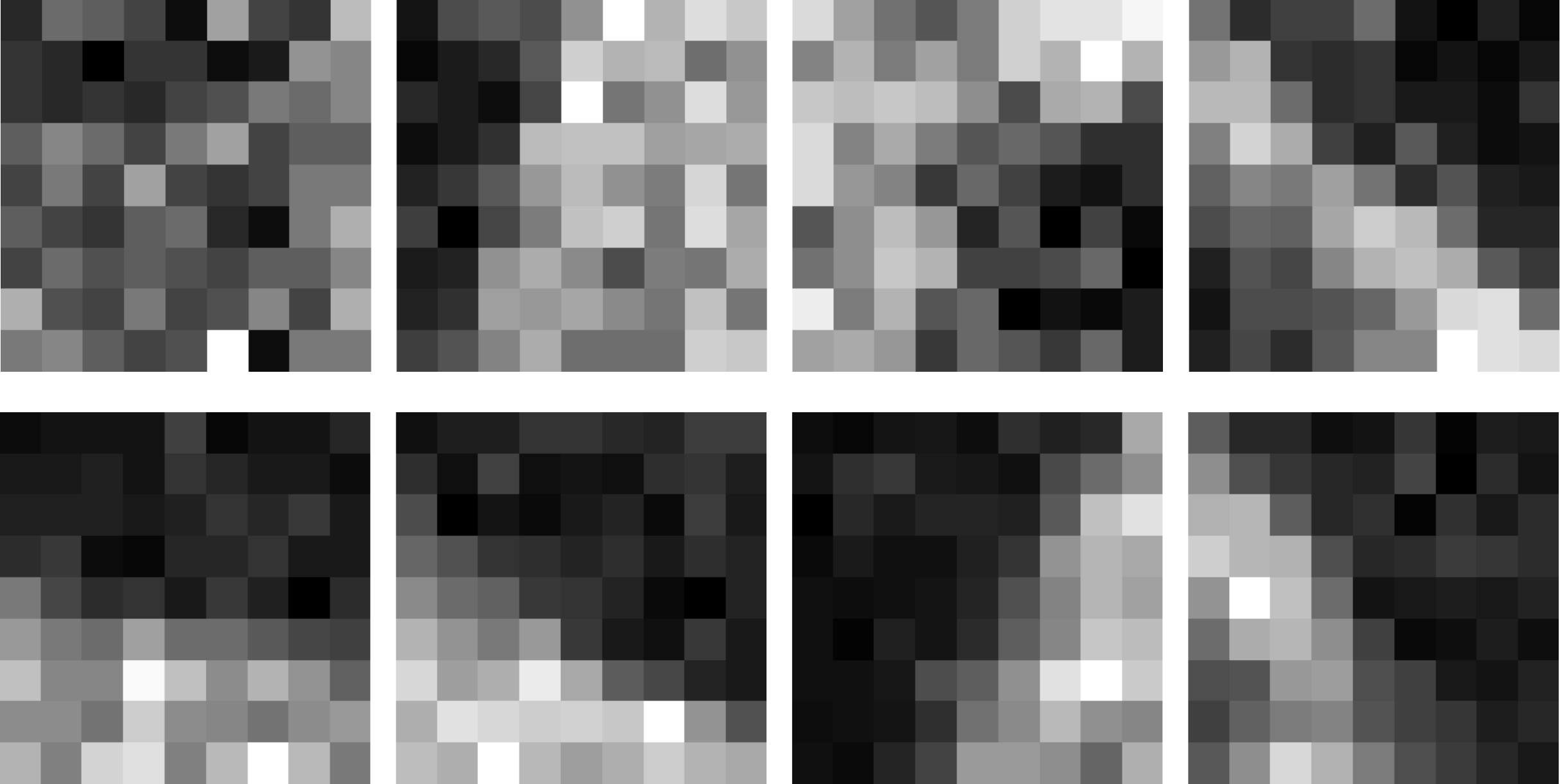}
    \caption{Examples of $9\times 9$ patches containing both blood vessels and background. The first and second rows of images show, respectively, patches containing 75\% and 50\% percent of blood vessel pixels.}
    \label{f:patchs_from_vessel_background_border}
\end{figure}

The values of $f=75\%$ and $f=50\%$ were used to verify whether the fraction of vessel pixels influences the classification performance. We generated 700 vessel patches for each value of $f$. For the background class, the same 700 patches generated in the previous experiment were used. Note that in this experiment, the \emph{vessel} class corresponds to patches that contain blood vessels, but also some background, whereas the patches in the \emph{background} class contain only background pixels. 
Figure~\ref{f:patchs_from_vessel_background_border} shows examples of extracted patches.

\subsubsection*{Selective removal of texture and intensity information}
 
Variations in contrast between vessel and background pixels and/or patterns that may exist in the arrangement of pixels in specific vessel regions can generate artifacts that influence the learning process of a neural network. Three variations of the original patch dataset $D$ were created by perturbing its original features to observe the impact of the loss of intensity and texture information about the samples. They involved intensity normalization and randomization of the pixel positions. Specifically:

{\bf Intensity normalization:}
A trivial way to classify a region of the image as a vessel or background is by comparing the intensities of the pixels. Therefore, it is interesting to measure the impact of removing intensity information about the patches. The perturbation of the intensities was achieved by applying normalization to zero mean and unit standard deviation, as defined by the equation

\begin{equation}\label{eq:z_score}
\tilde{x} = \frac{x - \mu}{\sigma}
\end{equation}
where $\mu$ and $\sigma$ are, respectively, the mean and standard deviation of the pixel intensities in a patch. This normalization removes the scale of the values as well as their variance, forcing the network to learn other features for patch classification. Note that the intensity normalization is relevant mostly for the experiment involving patches containing only vessel pixels or only background pixels, but it might also have an impact on patches containing border cues. The normalized patch dataset is henceforth referred to as $D_{\sm i}$.

{\bf Pixel shuffling:}
To investigate the importance of texture, a perturbation was applied to the spatial arrangement of the pixels by randomly shuffling their positions within each patch. Intensity values were not changed. The generated patch dataset $D_{\sm t}$ should be very challenging, as the structure of the vessels is completely lost. However, previous studies~\cite{zhang2017understanding} with natural images showed that a neural network can still successfully classify images with randomly shuffled pixels. 

{\bf Homogenization:}
The homogenized dataset is formed by the joint application of normalization and shuffling perturbations. The objective is to evaluate the performance under the most adverse condition: when information about local texture and the variation of pixel intensity is removed. This dataset is represented as $D_{\sm i,\sm t}$.

The four resulting datasets were used in the classification experiments with and without border information. Each set offers a distinct perspective on the information used by a neural network to differentiate between vessel and non-vessel pixels.

\subsubsection*{Model training and evaluation}

\begin{figure}[htbp]
    \centering
    \includegraphics[width=\columnwidth]{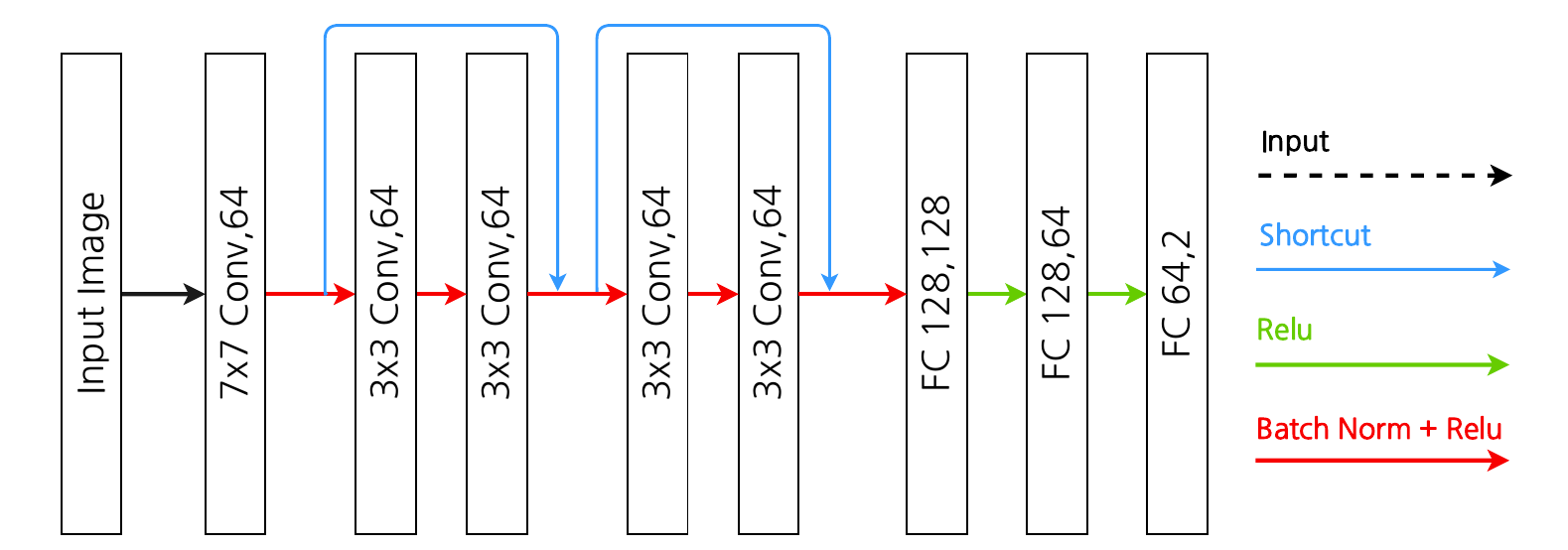}
    \caption{Neural network used in the patch classification experiments.}
    \label{f:model_classification}
\end{figure}

A simple classification model was trained and tested using the four previously mentioned datasets. Each dataset was randomly split into 80\% of the patches for training and 20\% for testing. The neural network used consists of two residual blocks~\cite{he2016deep} with 64 channels (illustrated in Figure~\ref{f:model_classification}), trained for 300 epochs with a learning rate of 0.001 and a cross-entropy loss function. A batch size of 9 was used, and no data augmentation was applied. The training runs were repeated five times with different splits of the dataset, and the average classification accuracy was calculated and used to evaluate the results.

\subsection{Isolating the influence of vessel shape}

It is a difficult task to isolate the shape cues used by neural networks for blood vessel segmentation. The style transfer approach developed by Geirhos et al.~\cite{geirhos2018imagenet} involves replacing the texture of whole images with randomly selected textures. The authors also experimented with replacing only the texture of the main object in the image, but this approach led to inconclusive results. Thus, their approach is not ideal for semantic segmentation. 

\begin{figure}[htbp]
    \centering
    \includegraphics[width=\columnwidth]{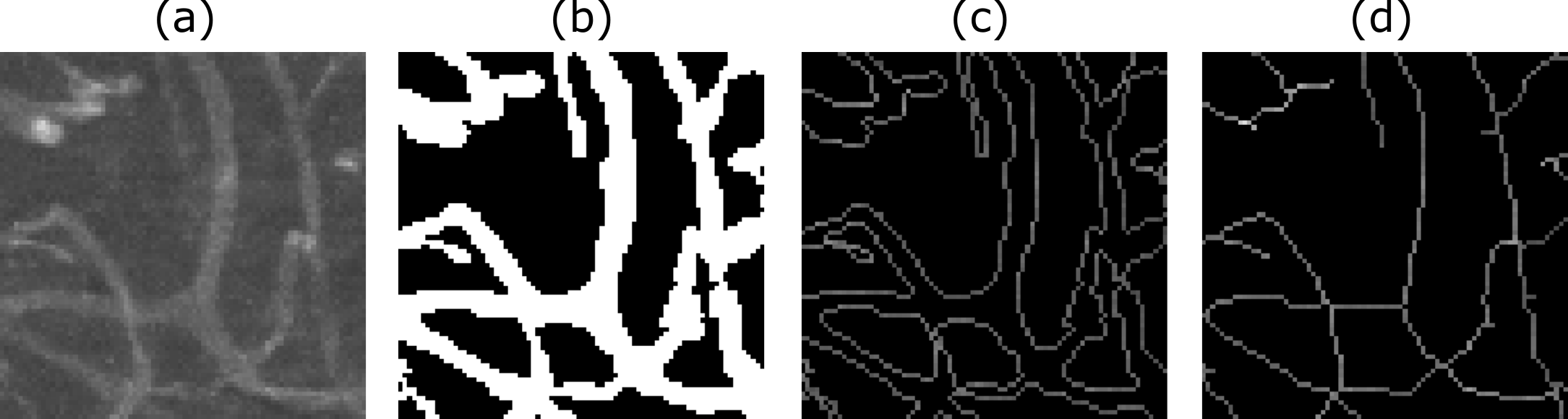}
    \caption{Example of images used for segmentation focused on vessel shape. (a) Original image. (b) Target image. (c) Sample where only the pixel intensities of the contour were kept. (d) Sample where only the pixel intensities of the centerlines were kept.}
    \label{f:skel_contour_examples}
\end{figure}

We developed an experiment to evaluate the ability of CNNs to segment blood vessels using only sparse shape information, removing internal vessel structure. The only cue of the model is the spatial location of the contour or centerline points of the blood vessels. Two new datasets were created for the experiment. The contour dataset $D_c$ was composed of samples containing only the contours of the blood vessels. The contours were obtained by eroding the segmentation masks of the vessels using a square structuring element and applying an XOR operation between the eroded image and the mask. The centerlines of the segmentation masks~\cite{lee1994building} were calculated and added to the result of the XOR operation to ensure that the erosion would not remove thin blood vessels from the image. To improve performance, instead of using binary contour images as input to the network, the original intensities of the pixels were assigned to the contours. An example is shown in Figure~\ref{f:skel_contour_examples}(c). 

A second, and likely more challenging, dataset $D_s$ was also created, comprising samples that contained only the centerlines of the vessels. As in the contour dataset, the original intensities of the pixels were kept at the centerline points. An example is shown in Figure~\ref{f:skel_contour_examples}(d). 

Samples from the datasets $D_c$ and $D_s$ were used as input to a CNN, and the targets were the segmentation masks of the \emph{original} samples from dataset $D$. For the dataset $D_c$, the network must learn to fill the interior of the contours to correctly predict the targets. Thus, inside the vessels it cannot use the texture and pixel intensities as cues for segmentation. Dataset $D_s$ represents an ill-posed problem. The network cannot correctly predict the caliber of the vessels from the centerline. As a best guess, it can learn the average caliber of the vessels and use it to predict the segmentation based on the centerline, or it might be able to use intensity cues to predict the local caliber. 

\subsubsection*{Model training and evaluation}

Datasets $D_c$ and $D_s$ were randomly split into 68, 12, and 20 samples for training, validation, and testing, respectively. The U-Net and W-Net models from Galdran et al.~\cite{galdran2022state} were trained on the data since they are simple architectures known to perform well on blood vessel segmentation tasks. The models were trained using the same code and procedures as the original authors, but with a larger number of epochs to ensure convergence. Specifically, the cross-entropy loss of the segmentation was optimized for 2500 epochs, and the model with the highest validation Dice score measured during training was used for testing. The training was carried out with a batch size of 4, a learning rate of 0.01, and with lightweight data augmentations consisting of a mixture of scaling, rotation, translation and vertical and horizontal flips.

\subsection{Measuring the amount of context required for good segmentation performance}
\label{s:rf_an}

Two approaches were developed to quantify the impact of context on segmentation. In the first approach, we selectively varied the parameters of the model associated with the RF size to evaluate the relationship between the RF size and the performance of the model. In the second approach, the input images were divided into patches and the model was trained to segment the vessels. These patches restricted the amount of context available, allowing us to quantify the patch size at which performance degrades due to insufficient context.

\begin{figure}[htbp]
    \centering
    \includegraphics[width=\columnwidth]{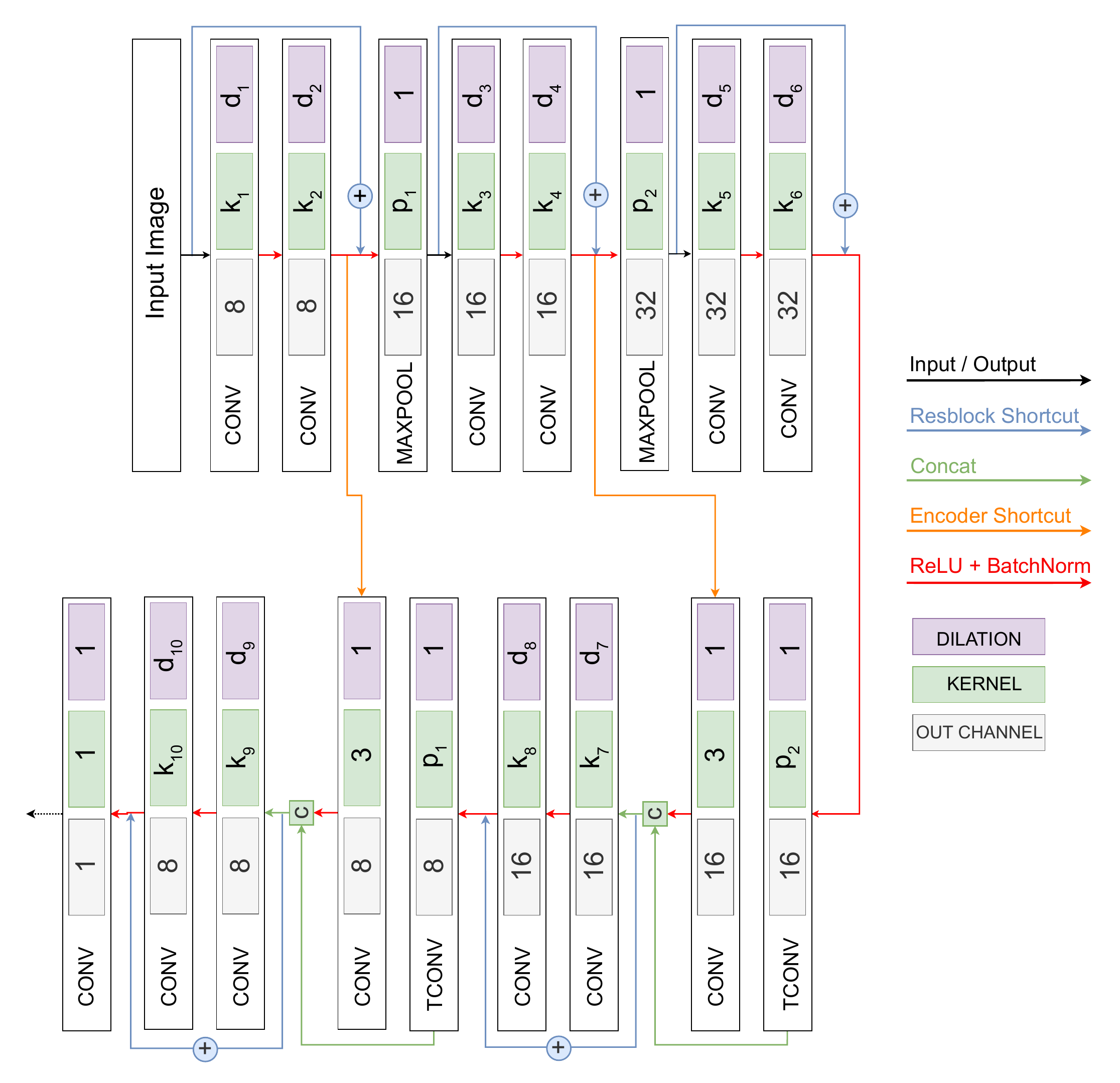}
    \caption{
    U-Net model used for the receptive field experiments. The parameters modified in the experiments are highlighted in green (kernel size) and purple (dilation ratio). For convolution layers, each kernel size $k_i$ was assigned a value from the set $\{1, 3, 5, 7\}$  and each dilation ratio $d_i$ was assigned a value from the set $\{2, 3, 7\}$. Pooling layers and their respective transposed convolution layers had $p_i=2$ when present, and $p_i=1$ when no pooling was applied.
    }
    \label{f:model_variations}
\end{figure}

For the first experiment, 160 different model configurations were trained on the VessMAP and DRIVE datasets. Different combinations of kernel size, dilation ratio, and downsampling amount were used. Figure~\ref{f:model_variations} shows the base architecture and the parameters changed for each layer. For instance, one model had all kernels with size 1, another model had the first kernel with a size of 1 while all others had a size of 3, and so on. The parameters used for all 160 models are listed in Appendix~\ref{app:models}. The model with a kernel size of 3 for all convolutions, a pool size of 2 for the max pooling layers, and a dilation ratio of 1 is the same as the one used in the previous section. This model is considered the \emph{base model} since it provides a strong baseline for vessel segmentation. For each model configuration, the model was trained for 1000 epochs using a learning rate of 0.01, a batch size of 4, and the cross-entropy loss. The model that achieved the highest validation Dice score during training was used for performance evaluation. Each training run was repeated three times to ensure statistical significance.

For the VessMAP dataset, the images were randomly split into training, validation, and test sets containing, respectively, 68, 12, and 20 images. For the DRIVE dataset, the official training set was randomly partitioned into 16 images for training and 4 for validation, and the official test set was used for testing. The data were augmented using random resizes, rotations, translations, and horizontal and vertical flips.  

After training the models, the size of the receptive field was measured. We considered two types of receptive fields: theoretical and effective. The theoretical receptive field (TRF) of an output pixel consists of all input pixels that can influence its value. It is calculated by finding the gradient of an output pixel with respect to all input pixels and keeping only the nonzero values. Since the TRF is a square, the length of one side was used as the TRF size. The ERF was measured using the methodology of~\cite{luo2016understanding}. Since the size can change depending on the output pixel, we averaged the calculated sizes for 100 randomly selected pixels from the test set of each dataset. The ERF ratio is defined as the ratio between the ERF and the TRF sizes. It measures the fraction of the TRF that is actually used by a CNN.

The second experiment consisted of dividing each sample of the VessMAP and DRIVE datasets into non-overlapping patches of size $W\times W$ and training the base model to segment the vessels. The model was trained using patches of sizes $[4, 8, 16, 32, 64, 128, 256, 512]$. The size 512 was not used in the VessMAP dataset because it is larger than the size of the images. We used the same training parameters, data splits, and augmentations as in the previous experiment. The only exception was the batch size. Since the amount of information contained in a patch varies according to $W$, a fixed batch size would lead to different amounts of information per batch for different values of $W$. Thus, the batch size was calculated for each patch size as 

\begin{equation}
    B = 4\left(\frac{H}{W}\right)^2,
\end{equation}
where $H$ is the size of the images in the dataset, all having a square aspect ratio. $H=256$ for the VessMAP dataset and $H=512$ for the DRIVE dataset. This ensured that the batches used for each patch size all contained the same number of pixels.

\section{Results}
\label{s:results}

The experiments were conducted on a Debian-based workstation equipped with an NVIDIA RTX 3080 GPU, an Intel Core i7-12700KF processor, and 32 GB of RAM. The implementation used Python 3.14 and PyTorch 2.10 (CUDA 13.0). The source code is publicly available at \url{https://github.com/chcomin/vascular-visual-cues}.

\subsection{Ablation of texture and intensity}
 
\begin{figure}
    \centering
    \includegraphics[width=\columnwidth]{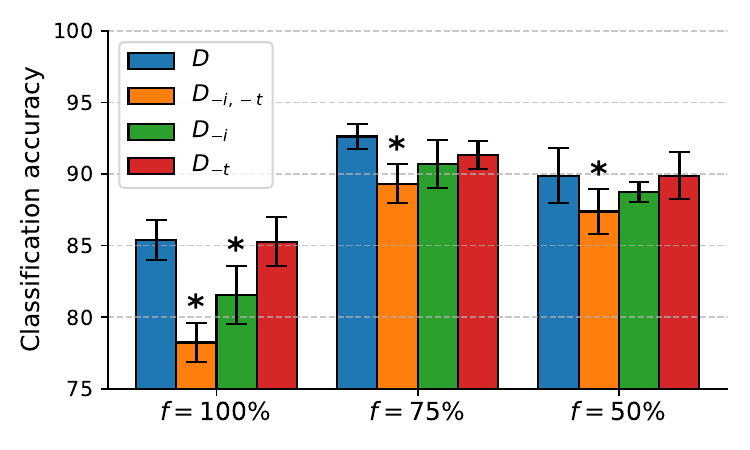}
    \caption{Average classification accuracy for different intensity and texture perturbations. The base dataset is represented as $D$. The $-i$ and $-t$ subscripts represent the removal of, respectively, intensity and texture information. Bar groups (100\%, 75\% and 50\%) represent different fractions of vessel pixels contained in the vessel class. Error bars show 95\% confidence intervals. Asterisks indicate a statistically significant decrease in accuracy compared to the original dataset ($p < 0.05$, one-tailed t-test).}
    \label{f:classification}
\end{figure}

The patch-based classification tests allow us to understand how the texture and intensity of the pixels help distinguish between vessels and the image background. Figure~\ref{f:classification} shows the classification accuracy of models trained on patches with and without border information, across the different perturbation datasets.

Starting with the experiment where vessel patches contain $f=100\%$ of vessel pixels, compared to $D$, the $D_{\sm i,\sm t}$ dataset shows an accuracy loss of 7.1\%. This result highlights that in images of blood vessels obtained by confocal microscopy, the spatial arrangement and intensity of the original pixels are relevant for the learning of the networks and, in turn, contribute to the correct classification of the pixels even without information from the global context of the image. Strikingly, $D_{\sm i,\sm t}$ still provides enough information for the neural network to classify patches based on other characteristics, achieving more than 75\% accuracy. The development of techniques to identify complex features used by the network besides pixel textures and intensities is an open problem to be analyzed in future studies.

Analyzing each image characteristic individually, we found that the loss of information with $D_{\sm t}$ is less than with $D_{\sm i}$. Therefore, modifying the spatial arrangement of the pixels had less impact on classification, resulting in an accuracy loss of only 0.1\%.

By including information from the blood vessel borders in the patches, we observed an improvement in the classification accuracy of the models for all datasets, as illustrated in Figure~\ref{f:classification}. Compared to the case $f=100\%$, we observed an accuracy gain of 7.3\% when the fraction of vessel pixels was $f=75\%$, and an increase of 4.5\% for $f=50\%$. This suggests that the geometry or contrast variations present in the vessel contours also contribute to the learning of CNNs. Pixel intensities still seem to be more relevant than texture, since $D_{\sm i}$ shows a greater loss of accuracy compared to $D_{\sm t}$.

The impact of the perturbations on the classification was smaller when including the vessel borders. For example, texture perturbations performed on patches where $f=50\%$ led to no impact on the classification performance. Similarly, the normalization of the pixel intensities also has a smaller impact when compared to the experiments without a vessel border. This suggests that as the informational content of the patches is enriched, the negative effect of the perturbations decreases. Thus, even in scenarios with greater image degradation, as the network gains access to information about the context in which the vessel is located, it may be able to compensate for part of the loss caused by the perturbations, contributing to a more robust classification.

It is important to mention that the benefit of adding border information appears to saturate, providing diminishing returns. This saturation is observed in the results of Figure~\ref{f:classification}, evidenced by the 2.6\% drop in the accuracy of the model when the number of vessel pixels decreases from 75\% to 50\%. This might be due to a larger proportion of background pixels in the patches. The network has a greater difficulty learning the characteristics of the blood vessels because it has less exposure to the pixels belonging to this class.

\subsection{Segmentation using shape cues}

Table~\ref{tab:shape_results} shows the results of the evaluation of CNN segmentation using only shape-related information, namely, through vessel contours $D_c$ or their centerlines $D_s$. The results are only shown for the VessMAP dataset since the values and conclusions were similar for the DRIVE dataset. The models did not achieve satisfactory performance.

The best performance was observed with the U-Net applied to $D_s$, reaching a Dice score of 57\%. For reference, the Dice score using the original samples containing full information was 90\%. Among the models trained with $D_c$, the W-Net model obtained a Dice score of 54.4\%, confirming that both approaches, contour and centerline, provide useful structural information for segmentation, yet are insufficient for accurate vessel segmentation. Furthermore, we observed a consistent pattern of high sensitivity and low specificity, particularly for the U-Net model applied to the $D_c$ dataset, which achieved a sensitivity of 93.4\% but a specificity of only 12\%. This indicates a strong tendency towards oversegmentation, resulting in many false positives.

\begin{table}
\caption{Performance of the U-Net and W-Net models on the test splits of the $D_c$ and $D_s$ datasets. Mean values and the respective errors of the mean are shown.}
\label{tab:shape_results}
%\centering
\begin{tabular}{llllll}
\toprule
Dataset & Model & Dice & Acc & Spec & Sens \\
        \midrule
        $D_c$ & U-Net & $44.8 \pm 0.4$ & $35.0 \pm 2.8$ & $12.0 \pm 1.9$ & $93.4 \pm 1.7$ \\
        $D_s$ & U-Net & $57.0 \pm 1.7$ & $67.6 \pm 2.3$ & $65.8 \pm 2.4$ & $72.0 \pm 2.9$ \\
        $D_c$ & W-Net & $54.4 \pm 0.8$ & $62.5 \pm 1.9$ & $55.9 \pm 1.3$ & $79.1 \pm 1.8$ \\
        $D_s$ & W-Net & $55.0 \pm 2.7$ & $59.1 \pm 2.0$ & $50.7 \pm 2.9$ & $80.6 \pm 2.0$ \\
\bottomrule
\end{tabular}
\end{table}

These results suggest that although vessel shape provides useful spatial information, it is insufficient for precise and reliable segmentation. Removing the internal vessel structure severely limited network performance. Thus, our findings reinforce the importance of combining both spatial (shape) and appearance (texture) features to develop more robust approaches for blood vessel segmentation.

\subsection{Linking segmentation performance with receptive field size}

\begin{figure*}[htbp]
    \centering
    \includegraphics[width=\textwidth]{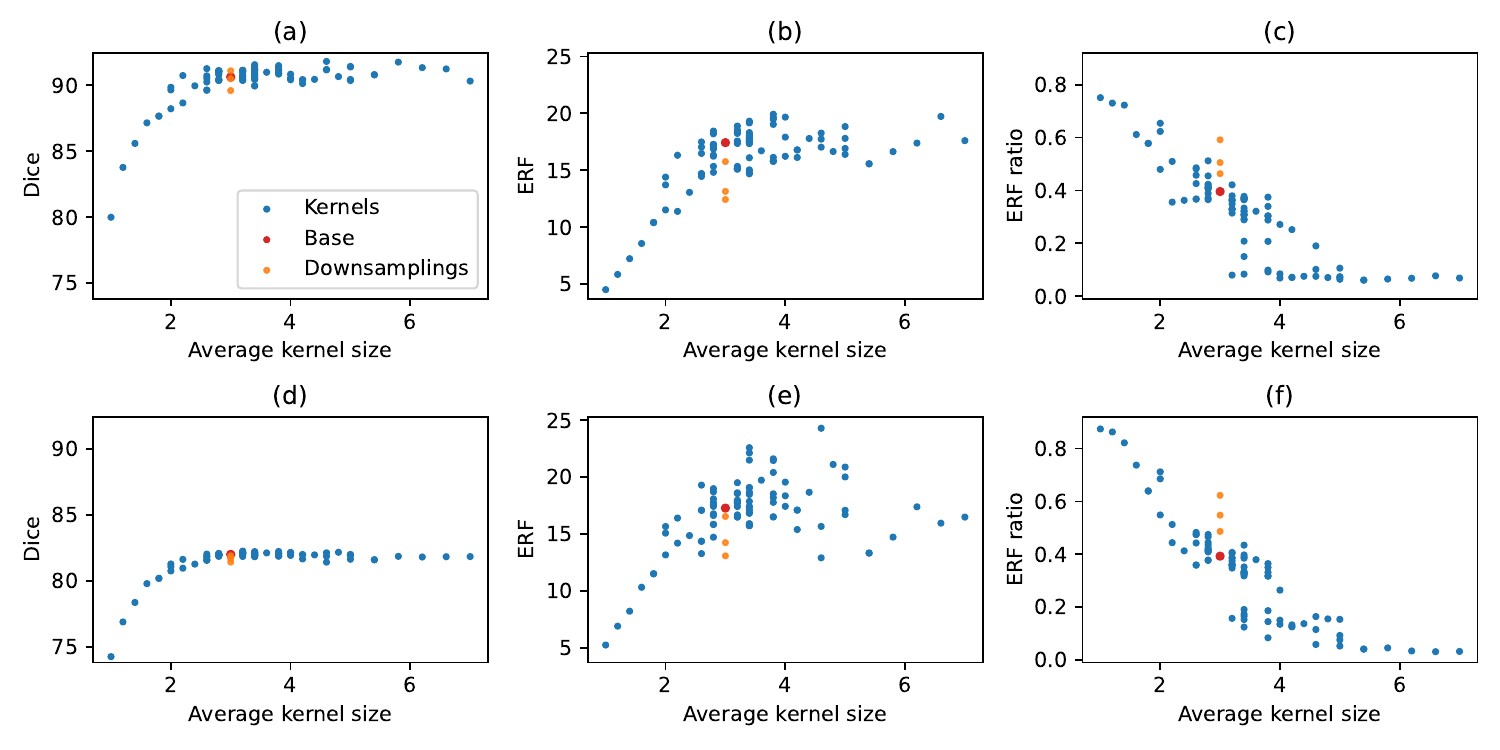}
    \caption{Segmentation performance and receptive field sizes measured for U-Net architectures having different kernel sizes. Plots (a), (b) and (c) are for the VessMAP dataset while plots (d), (e) and (f) are for the DRIVE dataset. The red dot shows the values of the base model while the orange dots show the results obtained when changing the downsampling layers.}
    \label{f:rf_kernel_variation}
\end{figure*}

Since the same number of kernels was maintained throughout the experiment and only their sizes were varied, the average kernel size of the model reflects the receptive field and also the number of trainable parameters. The receptive field increases with kernel size, which should benefit segmentation tasks. Increasing the kernel size also adds more trainable parameters and, in turn, increases the capacity of the model without altering the overall architecture. Figure~\ref{f:rf_kernel_variation} presents the results of the kernel variations on the VessMAP (Figures~\ref{f:rf_kernel_variation}(a), (b), and (c)) and DRIVE (Figures~\ref{f:rf_kernel_variation}(d), (e), and (f)) datasets.

Figures~\ref{f:rf_kernel_variation}(a) and (d) show that the performance of the model is positively correlated with the average kernel size. A similar trend is also observed for the relationship between the ERF and the average kernel size, as shown in Figures~\ref{f:rf_kernel_variation}(b) and (e). These results are consistent with the idea that larger kernels contribute to wider receptive fields, enabling a better capture of structures in the images. However, the improvements in the models' segmentation performance reached a plateau around an average kernel size of 3.

The size of the ERF is related to the practical utility of the captured information in reducing the loss and improving performance. Once the increase in kernel size no longer generates significant gains in segmentation, the gradients associated with regions further from the central area become less relevant, and the network ceases to substantially expand its ERF, even if the theoretical field allows for it.

The saturation observed in both analyses near an average kernel size of 3 reinforces a common trend in CNN literature: the use of the $3\times 3$ kernel as an ideal balance between complexity and efficiency in image segmentation models~\cite{simonyan2015very,he2016deep}.

The ratio between the ERF and the TRF, shown in Figures~\ref{f:rf_kernel_variation}(c) and (f), indicates how much of the receptive field is actually used by the CNN for segmentation. The ERF ratio decreases as the kernels become larger, demonstrating that the network utilizes a progressively smaller portion of the available TRF. Thus, the performance saturation is not a result of the physical limitation of the TRF. Regarding downsampling, removal of max pooling did not significantly reduce the Dice score, despite decreasing the ERF, suggesting that the network was able to compensate for the smaller contextual window.

\begin{figure*}[htbp]
    \centering
    \includegraphics[width=\textwidth]{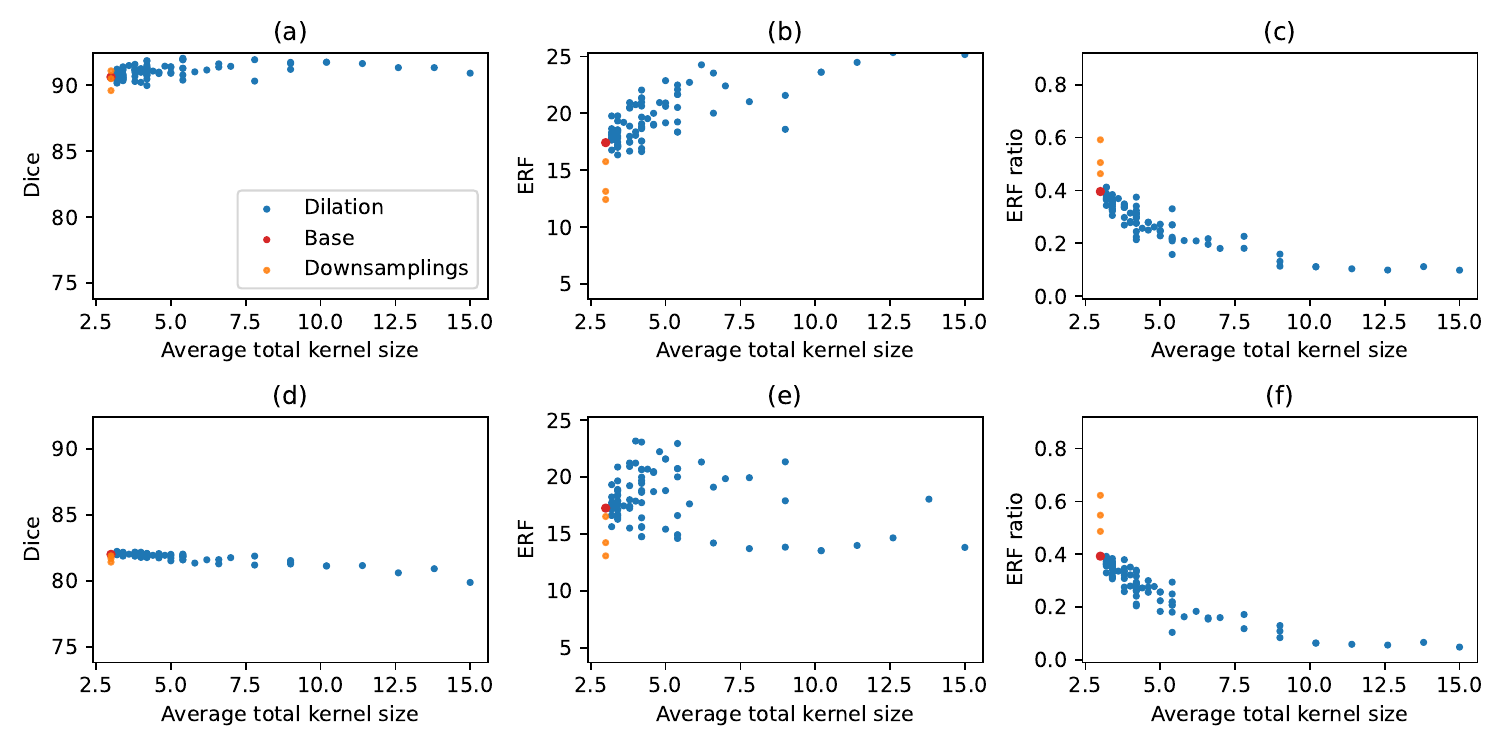}
    \caption{Segmentation performance and receptive field sizes measured for U-Net architectures having distinct dilation ratios. Plots (a), (b) and (c) are for the VessMAP dataset while plots (d), (e) and (f) are for the DRIVE dataset. The red dot shows the values of the base model while the orange dots show the results obtained when changing the downsampling layers.}
    \label{f:rf_dilation_variation}
\end{figure*}

The results of varying the dilation ratio can be seen in Figure~\ref{f:rf_dilation_variation}. The average total size of the kernels (kernel size times dilation rate) was used to analyze the impact of the increase in RF. We could not identify a clear correlation between kernel size and variations in the Dice score or the models' ERF. Although dilation can expand the model RF, there were no performance gains, indicating that the ERF of the base model was large enough. The only significant variation is a slight drop in the DICE score for the DRIVE dataset (Figure~\ref{f:rf_dilation_variation}(d)). Given that the vascular structures contained in the DRIVE dataset are small, the kernel gaps generated by the dilation might have led to the loss of relevant details of the vascular structures. 

Overall, the ERF tended towards a maximum value of around 20 pixels. The maximum receptive field achieved for a given dataset can guide the design of more efficient models, such that the receptive field's efficiency is maintained while complexity is reduced. The developed approach allows for identifying the point at which increasing complexity stops capturing relevant information, thereby avoiding unnecessary expansions that do not provide additional performance gains.

\begin{figure}[htbp]
    \centering
    \includegraphics[width=0.9\columnwidth]{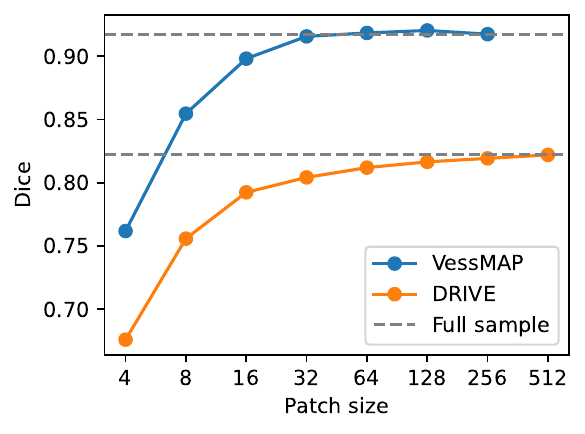}
    \caption{Dice score as a function of patch size for the VessMAP and DRIVE datasets.}
    \label{f:dice_vs_patch}
\end{figure}

To directly measure the RF required to achieve good segmentation performance, the patch extraction method described in Section~\ref{s:rf_an} was applied to the VessMAP and DRIVE datasets. Figure~\ref{f:dice_vs_patch} shows the Dice score as a function of patch size. Interestingly, performance saturates at size $32\times 32$ for the VessMAP dataset, that is, a $32\times 32$ context is sufficient to achieve maximum segmentation performance. This result is consistent with the maximum ERF observed in the previous experiment. However, performance does not saturate for the DRIVE dataset. The reason for this difference is unclear. We hypothesize that this is due to the DRIVE samples having a global structure, including an optic cup, large and small vessels at specific positions, and a well-defined region of interest (the retina). Further studies could investigate the mechanism underlying this difference. 

\section{Conclusion}
\label{s:conclusion}

We conducted a systematic investigation to disentangle the influence of texture, shape, and receptive field on the performance of CNNs for blood vessel segmentation. By treating fluorescence microscopy (VessMAP) and fundus photography (DRIVE) as representative case studies, we provided quantitative insights into the visual cues that guide decision-making in these specific imaging domains.

Our first set of experiments, conducted within a patch-based classification framework, suggests that for these modalities, pixel intensity is a more critical cue than local texture. The observation that models maintained performance above chance even after the removal of both intensity and texture indicates that the networks leverage other statistical features, the nature of which remains an open question for the community. Furthermore, the consistent improvement in accuracy when vessel borders were included highlights the significant role of local edge-based context.

The investigation into shape cues revealed that, within the parameters of our experimental setup, structural information alone is insufficient for reliable vessel reconstruction. The poor performance and tendency toward oversegmentation when using only contours or centerlines suggest that the evaluated CNNs rely heavily on the internal appearance and local statistics of vessels, rather than an inherent ability to extrapolate global vessel caliber from boundary geometry alone.

Regarding spatial context, our analysis showed that for the datasets studied, the Effective Receptive Field (ERF) used for segmentation tended to saturate at approximately 20 pixels. While the microscopy images (VessMAP) reached peak performance with purely local information, the fundus photographs (DRIVE) showed a modest benefit from larger contexts, likely due to the presence of global anatomical landmarks. This suggests that while most critical information is localized, the optimal receptive field is likely a function of the specific dataset's global structure.

Our finding that local context is often sufficient for accurate vessel delineation provides compelling evidence in the ongoing discussion regarding architecture choice. While architectures with global receptive fields, such as Transformers, offer theoretical advantages, our results are consistent with recent literature highlighting the continued efficacy of CNN-based approaches in vascular tasks~\cite{wittmann2025vesselfm,isensee2024nnu}, where local features may be of primary importance.

We acknowledge several limitations of this study. The results are empirical and derived from a specific subset of CNN architectures and two imaging datasets. Consequently, the findings regarding the ERF size and the relative importance of shape vs. texture should be viewed as observations within a controlled case study rather than universal properties of all neural networks. While these findings provide a strong baseline for microscopy and fundus images, future research is required to confirm these trends across a broader range of modalities and evolving architectures.

Ultimately, the insights derived from this study offer actionable pathways for new vascular segmentation models. The empirical saturation of the Effective Receptive Field at approximately 20 pixels suggests that lightweight, highly efficient CNNs can be purposefully tailored for local vessel delineation, avoiding the computational overhead of unnecessarily deep networks. Furthermore, these findings provide a roadmap for more effective integration with Explainable AI (XAI). Current saliency-based methods often fail to distinguish between different visual triggers, but future XAI frameworks could be designed to perform cue-specific attribution. Rather than providing generic heatmaps, these tools could decompose a model's decision into weighted contributions from intensity, texture, and shape. This would allow clinicians to determine if a model is identifying a vessel based on its anatomical structure or is simply being misled by local brightness artifacts, thereby directly addressing the black box concerns that hinder clinical trust. 

\section*{Acknowledgments}
Cesar H. Comin thanks FAPESP (grant no. 25/04800-9) for financial support. This study was financed in part by the Coordenação de Aperfeiçoamento de Pessoal de Nível Superior - Brasil (CAPES) - Finance Code 001.

\bibliography{references}

\appendix
\section*{Appendices} 

\renewcommand{\thesection}{\Alph{section}}

\renewcommand{\thefigure}{\thesection\arabic{figure}}
\renewcommand{\theHfigure}{\thesection.\arabic{figure}}
\setcounter{figure}{0}

\renewcommand{\thetable}{\thesection\arabic{table}}
\renewcommand{\theHtable}{\thesection.\arabic{table}}
\setcounter{table}{0}

\section{Model parameters used}
\label{app:models}

Section~\ref{s:rf_an} describes 160 models generated with different convolution kernel sizes, dilation ratios, and pooling layer strides. The specific kernel sizes and dilation ratios used for 157 of the 160 models are shown in Table~\ref{tab:parameters}. For these models, max pooling layers were always present. The remaining 3 models used a kernel size of 3 and dilation ratio of 1 for all layers, differing only by the presence of max pooling layers. Using the parameter names indicated in Figure~\ref{f:model_variations}:

\begin{itemize}
    \item Model ID 158: $p_1=2$ and $p_2=1$
    \item Model ID 159: $p_1=1$ and $p_2=2$
    \item Model ID 160: $p_1=1$ and $p_2=1$
\end{itemize}

\noindent $p_i=2$ indicates that the pooling layer was used, while $p_i=1$ means no pooling layer.

\begin{table*}
\caption{Kernel sizes and dilation ratios used for the models described in Section~\ref{s:rf_an}. Each ID indicates a different model. Using Figure~\ref{f:model_variations} as a reference, the \textit{kernels} columns show specific values used for the kernel size of the convolution layers in the format $k_1$:$k_2$:$k_3$:$k_4$:$k_5$:$k_6$:$k_7$:$k_8$:$k_9$:$k_{10}$. Similarly, the \textit{dilation ratios} columns show specific values in the format $d_1$:$d_2$:$d_3$:$d_4$:$d_5$:$d_6$:$d_7$:$d_8$:$d_9$:$d_{10}$. Models with ID 1-79 used a dilation ratio of 1 in all layers and models with ID 80-157 used a kernel size of 3 in all layers.}
\label{tab:parameters}
\centering
\renewcommand{\arraystretch}{1.2}
\small 
\begin{tabular}{
    l l               
    @{\hspace{1cm}}   
    l l               
    @{\hspace{1cm}}  
    l l               
    @{\hspace{1cm}}   
    l l               
}
\toprule
ID & Kernels & ID & Kernels & ID & Dilation ratios & ID & Dilation ratios \\
\midrule
1	& 3:3:3:3:3:3:3:3:3:3	& 41	& 5:5:5:5:5:3:3:3:3:3	& 80	& 2:1:1:1:1:1:1:1:1:1	& 119	& 3:3:3:3:3:1:1:1:1:1\\
2	& 1:3:3:3:3:3:3:3:3:3	& 42	& 5:5:5:5:5:5:3:3:3:3	& 81	& 1:2:1:1:1:1:1:1:1:1	& 120	& 3:3:3:3:3:3:1:1:1:1\\
3	& 3:1:3:3:3:3:3:3:3:3	& 43	& 5:5:5:5:5:5:5:3:3:3	& 82	& 1:1:2:1:1:1:1:1:1:1	& 121	& 3:3:3:3:3:3:3:1:1:1\\
4	& 3:3:1:3:3:3:3:3:3:3	& 44	& 5:5:5:5:5:5:5:5:3:3	& 83	& 1:1:1:2:1:1:1:1:1:1	& 122	& 3:3:3:3:3:3:3:3:1:1\\
5	& 3:3:3:1:3:3:3:3:3:3	& 45	& 5:5:5:5:5:5:5:5:5:3	& 84	& 1:1:1:1:2:1:1:1:1:1	& 123	& 3:3:3:3:3:3:3:3:3:1\\
6	& 3:3:3:3:1:3:3:3:3:3	& 46	& 5:5:5:5:5:5:5:5:5:5	& 85	& 1:1:1:1:1:2:1:1:1:1	& 124	& 3:3:3:3:3:3:3:3:3:3\\
7	& 3:3:3:3:3:1:3:3:3:3	& 47	& 5:3:5:3:5:3:5:3:5:3	& 86	& 1:1:1:1:1:1:2:1:1:1	& 125	& 3:1:3:1:3:1:3:1:3:1\\
8	& 3:3:3:3:3:3:1:3:3:3	& 48	& 3:5:3:5:3:5:3:5:3:5	& 87	& 1:1:1:1:1:1:1:2:1:1	& 126	& 1:3:1:3:1:3:1:3:1:3\\
9	& 3:3:3:3:3:3:3:1:3:3	& 49	& 3:3:5:5:3:3:3:3:3:3	& 88	& 1:1:1:1:1:1:1:1:2:1	& 127	& 1:1:3:3:1:1:1:1:1:1\\
10	& 3:3:3:3:3:3:3:3:1:3	& 50	& 3:3:3:3:5:5:3:3:3:3	& 89	& 1:1:1:1:1:1:1:1:1:2	& 128	& 1:1:1:1:3:3:1:1:1:1\\
11	& 3:3:3:3:3:3:3:3:3:1	& 51	& 3:3:3:3:3:3:5:5:3:3	& 90	& 2:2:1:1:1:1:1:1:1:1	& 129	& 1:1:1:1:1:1:3:3:1:1\\
12	& 1:1:3:3:3:3:3:3:3:3	& 52	& 3:3:3:3:3:3:3:3:5:5	& 91	& 2:2:2:1:1:1:1:1:1:1	& 130	& 1:1:1:1:1:1:1:1:3:3\\
13	& 1:1:1:3:3:3:3:3:3:3	& 53	& 3:3:3:3:3:3:5:5:5:5	& 92	& 2:2:2:2:1:1:1:1:1:1	& 131	& 1:1:1:1:1:1:3:3:3:3\\
14	& 1:1:1:1:3:3:3:3:3:3	& 54	& 7:3:3:3:3:3:3:3:3:3	& 93	& 2:2:2:2:2:1:1:1:1:1	& 132	& 7:1:1:1:1:1:1:1:1:1\\
15	& 1:1:1:1:1:3:3:3:3:3	& 55	& 3:7:3:3:3:3:3:3:3:3	& 94	& 2:2:2:2:2:2:1:1:1:1	& 133	& 1:7:1:1:1:1:1:1:1:1\\
16	& 1:1:1:1:1:1:3:3:3:3	& 56	& 3:3:7:3:3:3:3:3:3:3	& 95	& 2:2:2:2:2:2:2:1:1:1	& 134	& 1:1:7:1:1:1:1:1:1:1\\
17	& 1:1:1:1:1:1:1:3:3:3	& 57	& 3:3:3:7:3:3:3:3:3:3	& 96	& 2:2:2:2:2:2:2:2:1:1	& 135	& 1:1:1:7:1:1:1:1:1:1\\
18	& 1:1:1:1:1:1:1:1:3:3	& 58	& 3:3:3:3:7:3:3:3:3:3	& 97	& 2:2:2:2:2:2:2:2:2:1	& 136	& 1:1:1:1:7:1:1:1:1:1\\
19	& 1:1:1:1:1:1:1:1:1:3	& 59	& 3:3:3:3:3:7:3:3:3:3	& 98	& 2:2:2:2:2:2:2:2:2:2	& 137	& 1:1:1:1:1:7:1:1:1:1\\
20	& 1:1:1:1:1:1:1:1:1:1	& 60	& 3:3:3:3:3:3:7:3:3:3	& 99	& 2:1:2:1:2:1:2:1:2:1	& 138	& 1:1:1:1:1:1:7:1:1:1\\
21	& 1:3:1:3:1:3:1:3:1:3	& 61	& 3:3:3:3:3:3:3:7:3:3	& 100	& 1:2:1:2:1:2:1:2:1:2	& 139	& 1:1:1:1:1:1:1:7:1:1\\
22	& 3:1:3:1:3:1:3:1:3:1	& 62	& 3:3:3:3:3:3:3:3:7:3	& 101	& 1:1:2:2:1:1:1:1:1:1	& 140	& 1:1:1:1:1:1:1:1:7:1\\
23	& 3:3:1:1:3:3:3:3:3:3	& 63	& 3:3:3:3:3:3:3:3:3:7	& 102	& 1:1:1:1:2:2:1:1:1:1	& 141	& 1:1:1:1:1:1:1:1:1:7\\
24	& 3:3:3:3:1:1:3:3:3:3	& 64	& 7:7:3:3:3:3:3:3:3:3	& 103	& 1:1:1:1:1:1:2:2:1:1	& 142	& 7:7:1:1:1:1:1:1:1:1\\
25	& 3:3:3:3:3:3:1:1:3:3	& 65	& 7:7:7:3:3:3:3:3:3:3	& 104	& 1:1:1:1:1:1:1:1:2:2	& 143	& 7:7:7:1:1:1:1:1:1:1\\
26	& 3:3:3:3:3:3:3:3:1:1	& 66	& 7:7:7:7:3:3:3:3:3:3	& 105	& 1:1:1:1:1:1:2:2:2:2	& 144	& 7:7:7:7:1:1:1:1:1:1\\
27	& 3:3:3:3:3:3:1:1:1:1	& 67	& 7:7:7:7:7:3:3:3:3:3	& 106	& 3:1:1:1:1:1:1:1:1:1	& 145	& 7:7:7:7:7:1:1:1:1:1\\
28	& 5:3:3:3:3:3:3:3:3:3	& 68	& 7:7:7:7:7:7:3:3:3:3	& 107	& 1:3:1:1:1:1:1:1:1:1	& 146	& 7:7:7:7:7:7:1:1:1:1\\
29	& 3:5:3:3:3:3:3:3:3:3	& 69	& 7:7:7:7:7:7:7:3:3:3	& 108	& 1:1:3:1:1:1:1:1:1:1	& 147	& 7:7:7:7:7:7:7:1:1:1\\
30	& 3:3:5:3:3:3:3:3:3:3	& 70	& 7:7:7:7:7:7:7:7:3:3	& 109	& 1:1:1:3:1:1:1:1:1:1	& 148	& 7:7:7:7:7:7:7:7:1:1\\
31	& 3:3:3:5:3:3:3:3:3:3	& 71	& 7:7:7:7:7:7:7:7:7:3	& 110	& 1:1:1:1:3:1:1:1:1:1	& 149	& 7:7:7:7:7:7:7:7:7:1\\
32	& 3:3:3:3:5:3:3:3:3:3	& 72	& 7:7:7:7:7:7:7:7:7:7	& 111	& 1:1:1:1:1:3:1:1:1:1	& 150	& 7:7:7:7:7:7:7:7:7:7\\
33	& 3:3:3:3:3:5:3:3:3:3	& 73	& 7:3:7:3:7:3:7:3:7:3	& 112	& 1:1:1:1:1:1:3:1:1:1	& 151	& 7:1:7:1:7:1:7:1:7:1\\
34	& 3:3:3:3:3:3:5:3:3:3	& 74	& 3:7:3:7:3:7:3:7:3:7	& 113	& 1:1:1:1:1:1:1:3:1:1	& 152	& 1:7:1:7:1:7:1:7:1:7\\
35	& 3:3:3:3:3:3:3:5:3:3	& 75	& 3:3:7:7:3:3:3:3:3:3	& 114	& 1:1:1:1:1:1:1:1:3:1	& 153	& 1:1:7:7:1:1:1:1:1:1\\
36	& 3:3:3:3:3:3:3:3:5:3	& 76	& 3:3:3:3:7:7:3:3:3:3	& 115	& 1:1:1:1:1:1:1:1:1:3	& 154	& 1:1:1:1:7:7:1:1:1:1\\
37	& 3:3:3:3:3:3:3:3:3:5	& 77	& 3:3:3:3:3:3:7:7:3:3	& 116	& 3:3:1:1:1:1:1:1:1:1	& 155	& 1:1:1:1:1:1:7:7:1:1\\
38	& 5:5:3:3:3:3:3:3:3:3	& 78	& 3:3:3:3:3:3:3:3:7:7	& 117	& 3:3:3:1:1:1:1:1:1:1	& 156	& 1:1:1:1:1:1:1:1:7:7\\
39	& 5:5:5:3:3:3:3:3:3:3	& 79	& 3:3:3:3:3:3:7:7:7:7	& 118	& 3:3:3:3:1:1:1:1:1:1	& 157	& 1:1:1:1:1:1:7:7:7:7\\
40	& 5:5:5:5:3:3:3:3:3:3	& 	& 	& 	& 	& 	& \\
\bottomrule
\end{tabular}
\end{table*}

\end{document}